\documentclass[aoas,MSNbibl,nameyear,dvips]{arximspdf}
\usepackage{dcolumn}
\usepackage{graphicx}

\doi{10.1214/14-AOAS766} 
\volume{8}
\issue{4}
\pubyear{2014}
\firstpage{2223}
\lastpage{2246}
\docsubty{FLA}

\makeatletter
\newcolumntype{d}[1]{D{.}{.}{#1}}
\newcommand{\eqref}[1]{(\ref{#1})}
\makeatother

\begin{document}
\begin{frontmatter}

\title{Spatio-temporal modelling of extreme storms\thanksref{T1}}
\runtitle{Modelling extreme storms}
\thankstext{T1}{Supported by the project RACEWIN, which was
funded by the AXA research fund.}

\begin{aug}
\author[A]{\fnms{Theodoros}~\snm{Economou}\corref{}\ead[label=e1]{t.economou@ex.ac.uk}},
\author[A]{\fnms{David~B.}~\snm{Stephenson}\ead[label=e2]{d.b.s.stephenson@ex.ac.uk}}
\and
\author[A]{\fnms{Christopher~A.~T.}~\snm{Ferro}\ead[label=e3]{c.a.t.ferro@ex.ac.uk}}

\runauthor{T. Economou, D.~B. Stephenson and C. A. T. Ferro}

\affiliation{University of Exeter}

\address[A]{
College of Engineering, Mathematics\\
\quad and Physical Sciences\\
Exeter Climate Systems\\
University of Exeter\\
North Park Road\\
EX4 4QF, Exeter\\
United Kingdom\\
\printead{e1}\\
\phantom{E-mail:\ } \printead*{e2}\\
\phantom{E-mail:\ } \printead*{e3}}

\end{aug}
%

\received{\smonth{12} \syear{2012}}
\revised{\smonth{6} \syear{2014}}


\begin{abstract}
A flexible spatio-temporal model is implemented to analyse extreme
extra-tropical cyclones objectively identified over the Atlantic
and Europe in 6-hourly re-analyses from 1979--2009. Spatial variation in
the extremal properties of the cyclones is captured
using a 150 cell spatial regularisation, latitude as a covariate, and
spatial random effects. The North Atlantic Oscillation (NAO)
is also used as a covariate and is found to have a significant effect
on intensifying extremal storm behaviour, especially over
Northern Europe and the Iberian peninsula. Estimates of lower bounds on
minimum sea-level pressure are typically 10--50~hPa below the
minimum values observed for historical storms with largest differences
occurring when the NAO index is positive.

\end{abstract}

%
\begin{keyword}
\kwd{Bayesian hierarchical model}
\kwd{spatial random effects}
\kwd{natural hazards}
\kwd{extra-tropical cyclones}
\kwd{extreme value distribution}
\kwd{European windstorms}
\end{keyword}
\end{frontmatter}

\section{Introduction}\label{Intro}
Extreme North Atlantic and European extra-tropical cyclones are a major
source of risk for society.
These natural hazards cause much damage and insurance loss in Europe
due to extreme wind speeds/\break flooding. Recent
examples include the December 1999 windstorms Anatol and Lothar [\citet{Ulbrich:2001}], and windstorm Kyrill in 2007
which resulted in large losses across most of Europe. Important
scientific questions are as follows:
\begin{longlist}[1.]
\item[1.]
How extreme (intense) can extra-tropical cyclones become? Or, more
precisely, how much more extreme compared to the
most extreme values recorded in short series of historical
observations/analyses?
\item[2.]
How does the extreme behaviour vary spatially?
\item[3.]
How does the extreme behaviour vary in time due to modulation by
large-scale climate patterns?
\end{longlist}

We consider sea-level pressure (i.e., cyclone depth) as a measure of
cyclone intensity. Unfortunately, there are
no simple physical arguments for how deep an extra-tropical cyclone can
become. The most extreme events often deepen
\emph{explosively} with rapid decreases in central pressure, for
example, storms known as \emph{bombs} having pressure drops
of more than 24~hPa in 24 hours at 60~N. Explosive cyclogenesis depends
on many factors, for example, the
deepest recorded 20th century low of 913~hPa (the Braer cyclone of
January 1993) deepened 78~hPa in 24 hours
due to a combination of several factors such as available moisture and
stratospheric conditions [Odell et al. (\citeyear{Odell:2013})].
The unlikely possibility that such conditions could be maintained for 2
days gives a minimum value
of SLP of around $990-156=834$~hPa starting from a typical background
state of 990~hPa. It should also
be noted that SLP less than 650~hPa would correspond to mid-latitude
geostrophic wind speeds faster than
the speed of sound, which due to shock wave dissipation would be
impossible to maintain energetically.
In the absence of any more rigorous physical bounds, it is of interest
to estimate bounds empirically
using statistical approaches such as extreme value theory.

Modelling cyclones poses an interesting challenge: the events occur irregularly
in space and time with rates and magnitudes that are spatially
heterogeneous and nonstationary in time
(due to modulation by large-scale climate conditions). Furthermore, at
any one location, very few
extreme events are observed in short historical data sets. Here we
model extreme North Atlantic
cyclones using an extended version of the spatial point process model
for extremes from \citet{Cooley:2008}.
The extension involves the inclusion of temporal covariates, the
adaptation to
irregularly occurring (i.e., random occurrence rather than fixed
locations) extremes in space and the application
to extra-tropical cyclones.

\section{Background and data}
\subsection{Extreme extra-tropical cyclones}
There has been surprisingly little use of extreme value theory to investigate
extreme cyclones [see \citet{Katz:2010} for a discussion about the lack
of extreme value theory in climate science].
\citet{Lionello:2008} investigated changes in future cyclone climatology
over Europe using the Generalised Extreme Value (GEV)
distribution to model pressure minima. Return levels were calculated
over the whole North Atlantic domain without explicit
characterisation of spatial or temporal heterogeneity. \citet{Della-Marta:2009} and \citet{Della-MartaB:2009} used a Generalised
Pareto Distribution (GPD) model to analyse future changes in extreme
wind intensity. Three large nonoverlapping areas
were considered, however, there was no formal consideration of spatial
or temporal variation in the extremes.
\citet{Sienz:2010} used GPD models extending the work by \citet{Della-Marta:2009} to include temporal covariates such as the North Atlantic
Oscillation (NAO) and a linear trend but did not account for spatial
variability. \citet{Bonazzi:2012} used bivariate extreme value
copulas to model the spatial dependence in footprints of peak gust wind
speeds from a set of 135 damaging
European cyclones. However, this study did not explicitly model the
magnitude of many cyclones and so does not
answer the question about upper bounds on cyclone magnitudes.

\subsection{Brief review of spatial extreme models}
Davison, Padoan and Ribatet (\citeyear{Davison:2012}) identified three main classes of statistical models
for spatial extremes: Bayesian hierarchical models (BHM),
copula based models and max-stable process models. Although max-stable
processes explicitly characterise spatial dependence,
BHM can be more flexible and pragmatic by allowing for inclusion of
physical mechanisms in terms of covariates and random effects.
The major issue with BHM is the conditional independence assumption of
the extremes, whereas for max-stable processes it is model
implementation and flexibility. Copula models lie somewhere in between
since the dependence of the extremes is modelled by the
copula assuming that the marginal distributions are separable from this
dependency structure [\citet{Sang:2010}].

In this paper, we adapt BHM as the modelling framework mainly because
of their flexibility in allowing for (temporal) covariate
effects along with a versatile spatial dependency structure through
random effects.
BHM generally assume independence of the extremes for given values of
the covariates and random
effects (conditional independence), although they can be extended to
model spatial extremal dependence by including max-stable
processes [\citet{Reich:2012}]. For the application to extra-tropical
cyclones, we believe conditional independence to be a reasonable
working assumption. Much of the dependency between successive cyclones
has been shown to be induced by modulation of rates by
time-varying climate modes and so can be accounted for by including
appropriate covariates [\citet{Mailier:2006,Vitolo:2009}].

There has been recent interest in spatial BHM for extremes since their
introduction by \citet{Casson:1999}.
In \citet{Cooley:2007} and \citet{Cooley:2008}, a GPD and a point process model
are used to model extreme precipitation where the spatial dependence is
characterised by Gaussian
random effects in the formulation of model parameters but without
incorporating temporal nonstationarity.
\citet{Gaetan:2007}, \citet{Heaton:2009} and \citet{Sang:2009} allowed
temporal structure in BHM through time-varying
covariates where the conditional model is a GEV distribution.
\citet{Turkman:2010} used a similar model where the conditional model is
a GPD. In this paper, we use the computationally efficient
MCMC algorithm from \citet{Cooley:2008} based on recent work on Markov
random fields [\citet{Rue:2005}] and add temporal covariates,
to account for temporal trends and variations. We use the point process
model for extremes as the conditional model:
it utilises more of the data than GEV models and, unlike GPD models,
inference is invariant to the choice of threshold.

\subsection{Data}\label{Data}
\begin{figure}

\includegraphics{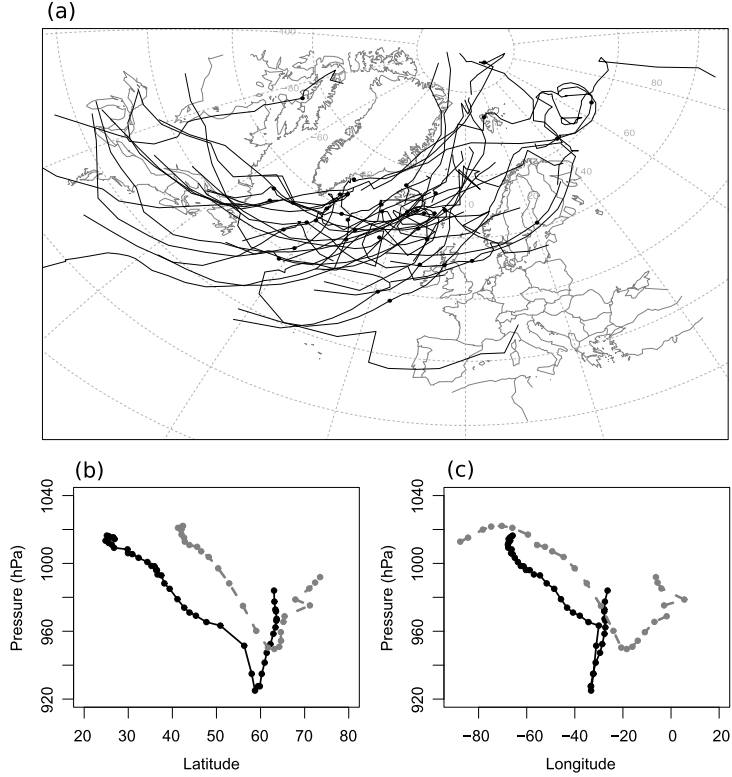}

\caption{\textup{(a)} Cyclone tracks for the October 1989 to March 1990 extended
winter. Only a subset of
tracks is plotted: ones where any 6-hourly MSLP value reached below 960~hPa.
Nadir positions are denoted with solid circles. \textup{(b)} Sea-level pressure
versus latitude and \textup{(c)} latitude for two of the cyclone tracks in \textup{(a)}.}
\label{Tracks}
\end{figure}

Objective feature-identification software [\citet{Hodges:1994}] was used
to extract cyclone tracks from
6-hourly National Center for Environmental Prediction Climate Forecast
System (NCEP-CFS) re-analysis data [\citet{Saha:2010}]
available over the period 1979--2009. Individual cyclone tracks are
identified by tracking local maxima in relative vorticity
just above the boundary layer (about 1.5~km above sea level). The
minimum sea-level pressure (MSLP) and its location
are recorded every 6 hours throughout the lifecycle of each cyclone. We
use sea-level pressure as a measure of cyclone intensity
mainly because this variable is well observed and has smooth variation
during the lifetime of a cyclone, unlike other possible
variables such as wind speed or vorticity. Figure~\ref{Tracks}(a) shows a
map of cyclone tracks defined by 6-hourly MSLP recordings
for a period with high cyclone activity. Only a subset of tracks is
plotted: ones where any 6-hourly MSLP value reached below
960~hPa. Typical damaging cyclones over Europe reach values in the range
940--970~hPa,
for instance, Anatol: 953~hPa [\citet{Ulbrich:2001}] and Kyrill: 962~hPa
[\citet{Mitchell:2007}], whereas the lowest ever recorded Braer
cyclone reached 913~hPa off the North--West of Scotland in January 1993
[\citet{Odell:2013}].

%

\begin{figure}

\includegraphics{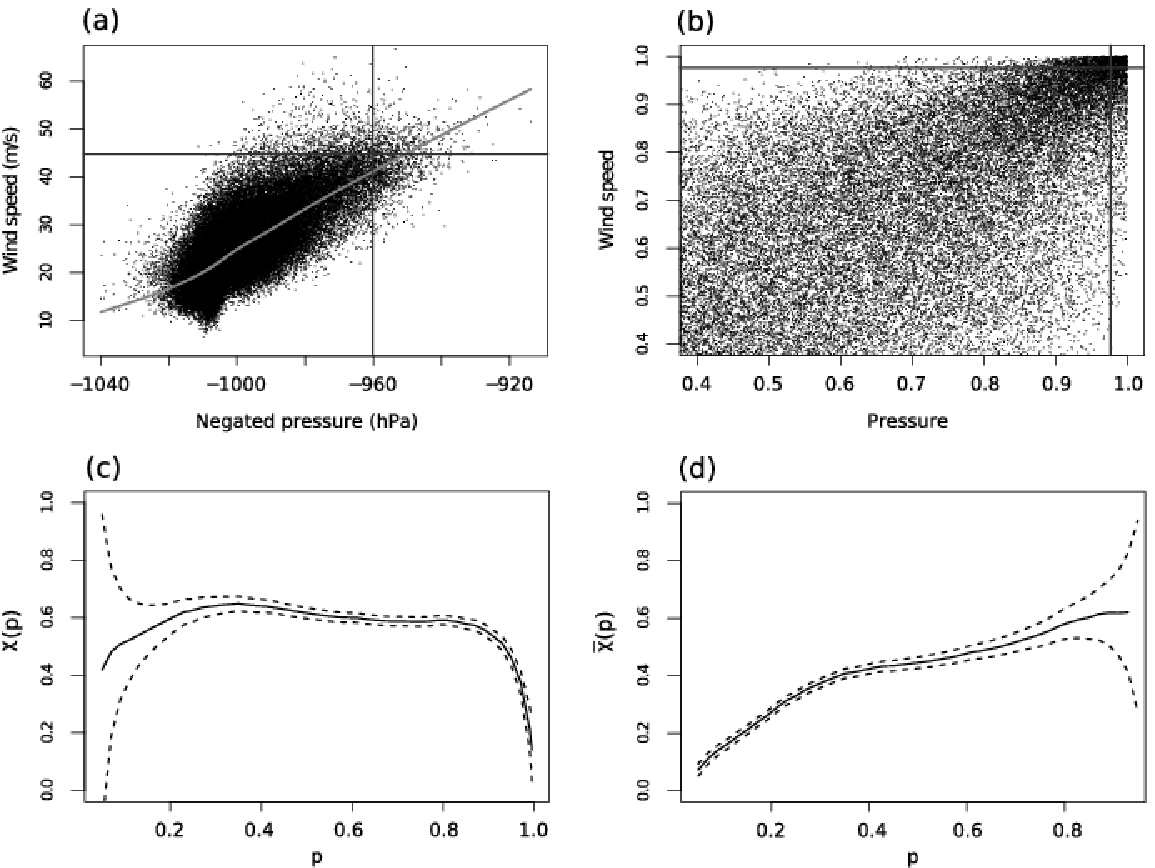}

\caption{\textup{(a)} Wind speed against negated sea-level pressure with an
associated loess fit (grey line). The intersecting lines
are the values $-$960~hPa and 45~m/s for pressure and wind speed,
respectively, representing the same high empirical quantile
for each variable. \textup{(b)} Empirical copula of wind speed and pressure
along with the associated quantile lines from \textup{(a)}.
\textup{(c)} The associated extremal dependency measure $\chi(p)$, and \textup{(d)} $\bar
{\chi}(p)$ vs $p$. The $95\%$ confidence intervals in \textup{(c)} and \textup{(d)}
are based on the Normal approximation to proportions and are calculated
as introduced in \citet{Coles:1999}.}
\label{wind_press}
\end{figure}

Although wind speed could also have been used, exploratory analysis
suggests that extreme MSLP
and maximum wind speed are strongly dependent, as to be expected from
simple balance arguments. Above the surface boundary layer outside
equatorial regions, centrifugal and Coriolis forces are approximately
balanced by the pressure
gradient force. Hence, wind speeds above the boundary layer in
extra-tropical cyclones are proportional to pressure gradients
(gradient wind balance). Surface pressure gradients in turn are
strongly related to the cyclone MSLP
since extra-tropical cyclones have similar synoptic spatial dimensions
(the so-called Rossby scale). Hence, from such simple
dynamical meteorology arguments, MSLP and maximum wind speeds are
expected to be extremally dependent
and so will convey similar information. Let random variables $W$ and
$Z$ denote maximum (6-hourly) wind speeds
at about 1.5~km above the surface (on the 925~hPa pressure surface) and
negated MSLP (obtained by multiplying MSLP by $-1$), respectively,
with associated 6-hourly recorded values $w_t$ and $z_t$.
Figure~\ref{wind_press}(a) shows a plot of $w_t$ against $z_t$: there is
strong positive association with the loess smoother indicating
a nearly linear relationship.
To better visualise extremal dependence, Figure~\ref{wind_press}(b) shows
the empirical copula obtained by producing a scatter plot of the
empirical probabilities $q^{(z)}_t = (\operatorname{rank}(z_t)-1)/(n-1)$ and
$q^{(w)}_t$ [\citet{Stephenson:2008}], where
$n$ is the total number of 6-hourly recorded values. This transforms
out the margins to uniform distributions
since $q^{(z)}_t$ and $q^{(w)}_t$ are estimates of the cumulative
distribution functions (CDFs) $F_{Z}(Z)$ and $F_{W}(Z)$.
Strong dependence of the extremes is evident from the convergence of
the points in the upper right-hand corner of the graph.

Figure~\ref{wind_press}(c) shows estimates of the extremal dependence
measure $\chi$ [\citet{Coles:1999}] defined as
$\chi=\lim_{p\to1} \chi(p)$, where $\chi(p)=\break \Pr(F_{Z}(Z) > p |
F_{W}(W) > p)$. As $p\to1$, $\chi(p)\to0$, implying asymptotic
independence, so we also show $\bar{\chi}(p)$, another measure of
strength of extremal dependence, in Figure~\ref{wind_press}(d).
The quantity $\bar{\chi}=\lim_{p\to1} \bar{\chi}(p)$ measures the
strength of extremal dependence within the class of
asymptotic independence. Since $\bar\chi(p)$ remains positive but does
not tend to 1, we conclude that there is a positive nonasymptotic
association at extremes of negated MSLP and maximum wind speed, so
either variable could potentially be used to investigate extremes
(see Appendix~\ref{Extremal_dependence} for details on $\chi$ and $\bar
{\chi}$).

Figure~\ref{Tracks}(b) and (c) show plots of MSLP
against latitude and longitude, respectively, for two particular cyclone
tracks in the 1989--1990 winter [Figure~\ref{Tracks}(a)].
The plots illustrate not only the tendency of intense cyclones to move
in a west-to-north direction but also the fact
that MSLP decreases (cyclone deepening) as the cyclone propagates in
space and time, to reach a minimum (which we assume
approximates the unobserved value of the cyclone nadir) before it
starts increasing again until the end of the life cycle.
Understanding the limiting strength of the nadirs is an important
aspect in the study of extra-tropical cyclones. However, the rate
of growth of cyclones depends on the large-scale atmospheric
environment that they pass from, so the pressure limit of cyclone nadirs
will vary with the spatial location of the cyclone. By only considering
the nadir from each track, we focus on a
fundamental limiting property of cyclones, namely how deep they can
get in general rather than how deep they can get in specific
spatial locations. In other words, we are interested in spatial
variation in cyclone intensity rather than
maximum local cyclone impact.

\begin{figure}

\includegraphics{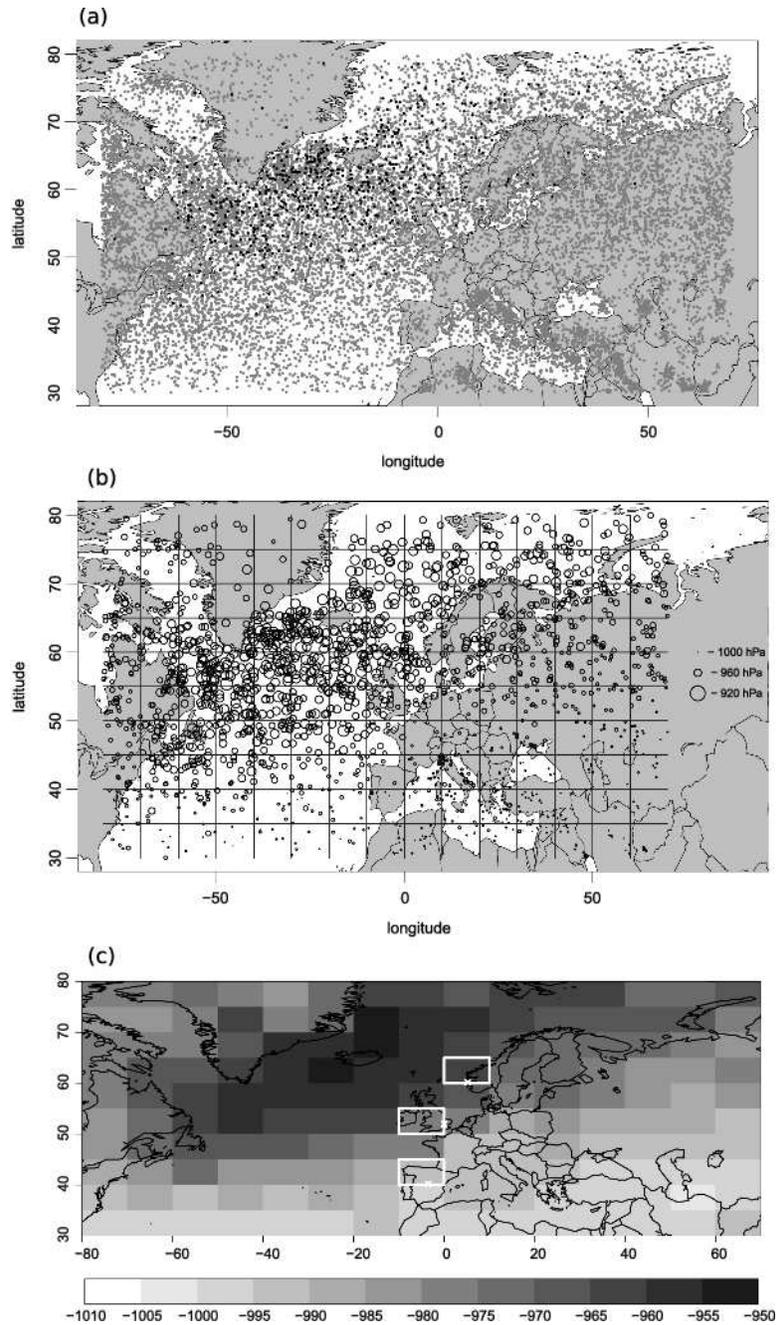}

\caption{\textup{(a)} Map of all cyclone nadirs: dots in black represent nadirs
deeper than 960~hPa, \textup{(b)} map of recorded $X(s,t)$ that are
greater than the threshold (90th empirical quantile) in each grid cell
and \textup{(c)} map of thresholds in each cell.}
\label{MAPS}
\end{figure}

The analysis of nadirs only, also helps to eliminate dependency between
successive 6-hourly MSLP measures and reduces the amount
of data from 313,557 6-hourly measurements to 17,230 nadirs. Figure~\ref{MAPS}(a) shows the (re-analysis) nadir from each track
in the Atlantic region where dots in black are nadirs with sea-level
pressure lower than 960~hPa. However, a single value for the
threshold defining the extremes is not appropriate and the definition
of extremeness should vary spatially. For example, a damaging
cyclone in the Mediterranean is likely to be considered a weak one over
Scandinavia.

Note that we use cyclone tracks from a reanalysis data set mainly
because generally cyclone track observations for the
extra-tropics are not readily available. However, reanalysis data are
output from climate models with assimilated historical
observational data. There is much smoothing/\break interpolation of the
observational data when creating a reanalysis data set, so the
interpretation of any results obtained here is conditional on the
effects of such smoothing.\

\section{Model specification and model fitting}\label{Application}
\subsection{Spatial discretisation}\label{spatial_discretisation}
Conventional Bayesian spatial models generally rely on the assumption
that data are either gridded or
they come from fixed locations in space [see \citet{Banerjee:2004}],
where one or more observations are available at each location.
Extreme nadirs, however, behave like a spatial marked point process
where both location of occurrence and
magnitude are random. To utilise such Bayesian models, we propose for
simplicity to discretise space by imposing a finite grid
and to consider the minimum possible size $\Delta$ for each grid cell,
to ensure that enough data are available for estimation
in each cell. Inference should not be sensitive to the choice of grid
spacing, provided it is fine enough (in the limit
$\Delta\rightarrow0$ one should obtain the original marked point
process). Sensitivity analysis for $\Delta$ is an important
part of the concept (see Section~\ref{Sensitivity}).

For spatial marked point processes, estimation is only possible after
making assumptions about spatial (and temporal) structure.
The assumption made by discretising is that, conditional on a
cell-specific random effect and possible covariates, the extreme events
(nadirs) within each cell come from the same distribution. The
cell-specific random effects are spatially dependent to allow for
correlation between events in neighbouring cells. We also assume that
the events can occur anywhere within the cells, with equal
probability. Importantly, redefining space into discrete grid cells
also provides a way of defining extremes in space: as values
below a cell-varying threshold or as the $r=1,2,\ldots$ largest values,
in fixed time periods.

More generally, conditional on a given spatial or spatio-temporal
dependence structure between cells, nadirs are modelled using an
appropriate extreme value model. This is a hierarchical model where at
the top of the hierarchy, random effects and covariates define
a spatio-temporal process which modulates the process, giving rise to
extreme nadirs.

\subsection{Spatial grid}\label{Spatial_Grid}
Conventionally, extreme value modelling is applied to the upper tails
so the nadirs are negated to obtain variable $X(s,t)$, where $s$
refers to the grid cell and $t$ refers to time. We may think of
$X(s,t)$ as the depth of a cyclone so that high values of $X(s,t)$
correspond to low values of MSLP. We divided the domain in Figure~\ref{MAPS}(a) into $N=150$ $5^{\circ}\times10^{\circ}$ grid cells.
The threshold $u(s)$ in each cell was defined as the empirical 90th
quantile of $X(s,t)$. We performed exploratory threshold analysis
using mean residual life plots [\citet{Coles:2001}], ensuring that the
90th empirical quantile was an appropriate threshold choice.
Figure~\ref{MAPS}(b) shows the map of the extremes (1736 nadirs) and
Figure~\ref{MAPS}(c) shows the map of $u(s)$. Note that in
Figure~\ref{MAPS}(c), three cells are highlighted: cells containing
coordinates ($5.2^\circ$E, $60.2^\circ$N), ($0^\circ$E, $5^\circ$N)
and ($3.5^\circ$W, $40.2^\circ$N) marked in white crosses. These
coordinates relate to the cities of Bergen, London and Madrid, respectively,
and will be used throughout the paper for illustration of results, as
they adequately span Europe in terms of latitude.

\subsection{Model specification}\label{model_spec}
To model the depth $X$ of negated nadirs, we consider the point process
model for extremes [\citet{Coles:2001}]---conditional on spatial random effects and temporal covariates. For some
high threshold $u$ of $X$, this model is parametrised
in terms of the location, scale and shape parameters of the GEV
distribution, namely, $\mu$, $\sigma$ and $\xi$
(see Appendix~\ref{Appendix}). We use the notation $X \sim\operatorname{PP}(\mu
,\sigma,\xi,u)$. Introducing spatial and temporal
variation, let $X(s,t)$ be the depth in grid cell $s\in\mathbb{S}$ at
time $t\in\mathbb{T}$, where $\mathbb{S}$ and $\mathbb{T}$
are the space and time domains, each a fixed subset of 2-dimensional
and 1-dimensional Euclidean space, respectively.
Extending the approach of \citet{Cooley:2008}, we model the $X(s,t)$ in
the following way:
%
\begin{eqnarray}
X(s,t)|\theta^\psi(s),\beta_2(s) &\sim& \operatorname{PP}
\bigl(\mu(s,t),\sigma (s,t),\xi(s),u(s)\bigr), \label{cyclonesModel}
\\
\mu(s,t) &=& \beta_0^{\mu} + \beta_1^{\mu}z_1(t)
+ \beta_2(s)z_2(t) + \theta^{\mu}(s),
\label{mu}
\\
\log\bigl(\sigma(s,t)\bigr) &=& \beta_0^{\sigma} +
\beta_1^{\sigma}z_1(t) + \theta ^{\sigma}(s),
\label{sig}
\\
\xi(s) &=& \beta_0^{\xi} + \theta^{\xi}(s)
\label{xi}
\end{eqnarray}
for $\psi=\mu,\sigma,\xi$. Defining vectors $\mathbf{U}^\psi=
( U^\psi(1),\ldots,U^\psi(N)  )'$ for $\psi=\mu,\sigma,\xi$,
$\mathbf{U}(s)= (U^\mu(s),U^\sigma(s),U^\xi(s) )'$ and
$\mathbf{U}= (\mathbf{U}^{\mu},\mathbf{U}^{\sigma
},\mathbf{U}^{\xi} )'$,
the spatial level of the model is as follows:
%
\begin{eqnarray}
\bigl(\theta^{\mu}(s), \theta^{\sigma}(s), \theta^{\xi}(s)
\bigr)'|\mathbf{U}(s) &\sim& \mathrm{N} \bigl(\mathbf{U}(s),
\operatorname {diag}(\bolds{\tau})^{-1} \bigr),\label{Thetacyclones}
\\
\mathbf{U}=\bigl(\mathbf{U}^{\mu}, \mathbf{U}^{\sigma},
\mathbf{U}^{\xi}\bigr)' &\sim& \mathrm{N} \bigl(\mathbf{0},
\bolds {\Omega}^{-1} \bigr),
\\
\beta_2(s) &\sim& \mathrm{N} \bigl(\nu,\phi^2 \bigr),
\end{eqnarray}
where $z_1$ is the latitude of the occurrence and $z_2$ is the North
Atlantic Oscillation (NAO) value (see Section~\ref{CovSel} about
covariate selection).
The spatial random effects $\theta^{\mu}(s)$, $\theta^{\sigma}(s)$ and
$\theta^{\xi}(s)$ define spatial variability in
$\mu$, $\log(\sigma)$ and $\xi$ across the cells, after allowing for
covariates. The $r$-year return level, that is, the $(1-1/r)$th
quantile of $X(s,t)$ in cell $s$ and time $t$, is given by
%
\begin{equation}
\label{RL} X_{1-1/r}(s,t)=\mu(s,t) + \frac{\sigma(s,t)}{\xi(s)}\bigl( \bigl(-
\log (1-1/r)\bigr)^{-\xi(s)} -1\bigr).
\end{equation}

As in \citet{Cooley:2008}, vectors $\mathbf{U}^{\psi}$ are modelled
jointly using a separable formulation [\citet{Banerjee:2004}, Chapter~7],
so that the precision matrix is $\bolds{\Omega} = \mathbf{T}
\otimes \mathbf{W}$. The matrix $\mathbf{T}$ is an
unknown $3\times3$ positive
definite symmetric matrix and $\mathbf{W}$ is an $N\times N$
proximity matrix defining spatial proximity between the $N$ cells.
Therefore, the dimension of $\bolds{\Omega}$ is $3N\times3N$.
Here, spatial proximity is based on nearest neighbours so that
off-diagonal elements
of $\mathbf{W}$ are $w_{i,j}=-1$ if cells $i$ and $j$ are adjacent
and $w_{i,j}=0$ otherwise, whereas diagonal elements
$w_{i,i}=-\sum_{i\neq j} w_{i,j}$ [see \citet{Bailey:1995}, pages
261--262 for other examples of proximity measures].

Each vector $\mathbf{U}^{\mu}, \mathbf{U}^{\sigma}, \mathbf
{U}^{\xi}$ is modelled by an Intrinsic AutoRegressive (IAR) spatial
model [\citet{Banerjee:2004}]. The IAR model uses the proximity matrix
and a single unknown parameter to control the spatial dependency structure
(see Appendix~\ref{IAR}). Here, there are three such parameters for
each of $\mathbf{U}^{\mu},
\mathbf{U}^{\sigma}, \mathbf{U}^{\xi}$ and they are found in
the diagonal of $\mathbf{T}$. (Note that the value of $\bolds
{\tau}$
is conventionally fixed beforehand to avoid nonidentifiability between
$\bolds{\tau}$ and the diagonal of $\mathbf{T}$
[\citet{Banerjee:2004}].) Dependence between $\mathbf{U}^{\mu},
\mathbf{U}^{\sigma}, \mathbf{U}^{\xi}$ is modelled using 3
parameters, the off-diagonals of $\mathbf{T}$, each controlling the
strength of dependence. Allowing explicitly for this dependence
can aid the MCMC estimation discussed in Section~\ref{MCMC}, in terms
of convergence to the posterior and also mixing of the MCMC samples.

The NAO parameter $\beta_2(s)$ is spatially variable but in an
unstructured way. This ensures that $\beta_2(s)$ share information to aid
estimation in cells with few events but less so compared to using a
structured (IAR) spatial prior. Parameter $\nu^{\psi}_k$ reflects the
overall NAO effect on $\mu(s,t)$.

We complete the model specification by defining the prior distributions
of the hyperparameters.
The intercepts $\beta^{\mu}_0,\beta^{\sigma}_0,\beta^{\xi}_0$ were
given Gaussian priors with large
variance, and means $(\bar{\mu},\log\bar{\sigma},\bar{\xi})$,
calculated as means of independent maximum likelihood fits of point
process models in each cell. For parameters $\beta_1^{\mu},\beta
_1^{\sigma},\nu$, we assumed a flat Gaussian
prior with zero mean and large variance.
The prior distribution $\pi(\cdot)$ for $\phi^\psi_k$ is chosen so that $\pi
(\phi^\psi_k)\propto1/\phi^\psi_k$ [\citet{Gelman:2004}, Chapter~3], whereas
for $\mathbf{T}$ and $\mathbf{P}$ we use a Wishart prior with 3
degrees of freedom (uninformative) and a mean that relates
to the variability of $\mu$, $\sigma$ and $\xi$ across cells (see
Section~\ref{MCMC}).

\subsection{Covariate selection}\label{CovSel}
This was performed by adding explanatory variables to a ``null'' model:
the model in (\ref{cyclonesModel})--(\ref{xi}) without $z_1$ or $z_2$.
Models were compared using the Deviance Information Criterion (DIC), a
model selection criterion for Bayesian models [\citet{Spiegelhalter:2002}]
and by investigating whether posterior distributions of associated
parameters are centred at zero with relatively large variance.

The model in (\ref{cyclonesModel})--(\ref{xi}) was first implemented
with the addition of latitude, longitude, latitude squared,
longitude squared and an interaction term between longitude and
latitude as covariates in both $\mu(s,t)$ and $\log(\sigma(s,t))$.
This allows for large-scale spatial trends, leaving the local spatial
dependence to the random effects. It also relaxes the assumption
of complete spatial randomness of extreme events within a cell, both in
terms of occurrence and intensity. In principle, nonparametric
surfaces can also be considered for smoothing large-scale spatial
trends [see \citet{Davison:2012}, page 173 for references], but this was
not deemed necessary here. To quantify the effect of large-scale
climate patterns, two climate indices were also considered as covariates:
the North Atlantic Oscillation (NAO) and the East Atlantic Pattern
(EAP), both of which have been shown to be influential for extra-tropical
cyclones [\citet{Mailier:2006,Seierstad:2007,Pinto:2009,Nissen:2010}].
No covariates were considered for the shape parameter $\xi(s)$ since
this is a particularly difficult parameter to estimate, however, it was
allowed to vary between cells. Out of all possible covariate
combinations, the lowest DIC value occurred for the particular model
formulation in (\ref{cyclonesModel})--(\ref{xi}). The posterior
distributions of ``insignificant'' parameters (e.g., ones relating to
longitude) had means and medians very close to zero.

\begin{figure}

\includegraphics{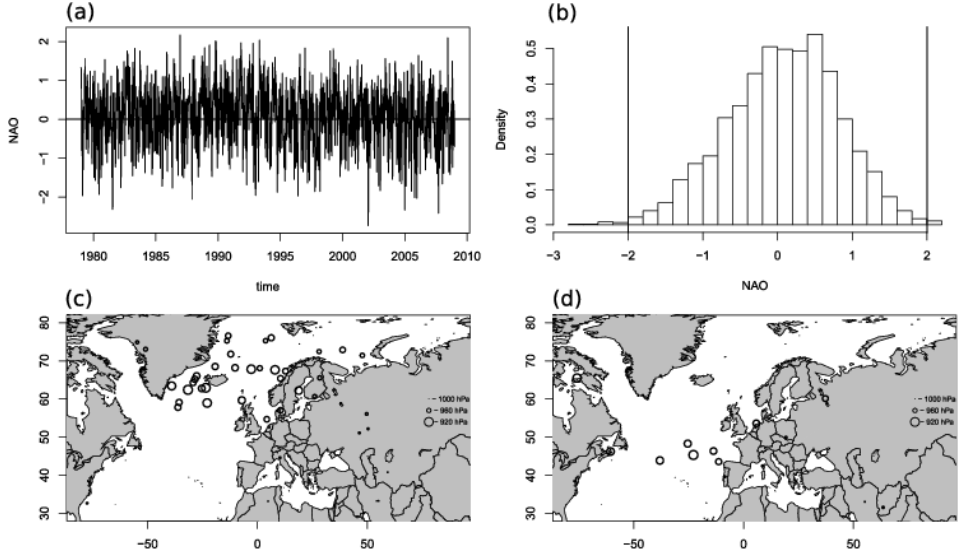}

\caption{\textup{(a)} Time series of NAO defined as a 5-day average of daily
NAO, \textup{(b)} histogram of NAO along with vertical lines marking the values
$-2$ and 2, \textup{(c)} occurrences of recorded nadirs where the associated NAO
value was greater than 2 and \textup{(d)} less than $-2$.}
\label{NAOplots}
\end{figure}

It is well known that the NAO has influence on the development of
extra-tropical cyclones [\citet{Pinto:2009}]. By definition, the
NAO index is standardised to have zero mean and unit variance, and here
it was defined as 5-day nonoverlapping
averages from 1979--2009. Figure~\ref{NAOplots}(a) shows the time series
of NAO and Figure~\ref{NAOplots}(b) shows the histogram of NAO
where the values of 2 and $-2$ are marked, as we consider these as high
and low NAO threshold values throughout the rest of this paper.
Figure~\ref{NAOplots}(c) and (d) show extreme values of
$X(s,t)$ for which $\operatorname{NAO} \geq2$ and $\operatorname{NAO} \leq-2$, respectively.
There is a clear North--South pattern in the Central Atlantic, implying
NAO has a notable effect on extreme cyclones.

\subsection{Estimation by Markov chain Monte Carlo}\label{MCMC}
For all $\psi=\mu,\sigma,\xi$, random effects $\theta^{\psi}(s)$ and
$\beta_2(s)$, and parameters $\beta_1^{\mu}$ and
$\beta_1^{\sigma}$, were sampled by Metropolis--Hastings, specifically
using a random walk sampler. The intercepts
$\beta^{\psi}_0$ were sampled from their full conditionals using Gibbs
sampling, by treating them as intercepts in the mean
for each $\theta^{\psi}(s)$. Samples of $(\mathbf{U}^{\mu},
\mathbf{U}^{\sigma}, \mathbf{U}^{\xi})'$ and
$\mathbf{T}$ were drawn using Gibbs sampling, utilising the
specific techniques in \citet{Cooley:2008}. Both
$\nu$ and $\phi^2$ were sampled from their full conditionals: Gaussian
and scaled inverse-$\chi^2$, respectively.

Note that when the IAR model is used as a prior it is improper: the
density does not integrate to 1. So, to make
the intercept terms $\beta^{\psi}_0$ identifiable, the rows of
$\mathbf{W}$ must sum to zero. This in turn imposes
the restriction that $\sum_s U^{\psi}(s)=0$.

The parameter $\bolds{\tau}$ was set to $(0.1, 10, 100)'$. These
values were chosen
by fitting independent point process models in each cell and
investigating the level of variability between cells for
$\mu(s)$, $\log(\sigma(s))$ and $\xi(s)$, not only to reflect the
difference in scale for the three parameters
but also to make sure that most of the variability is modelled by the
random effects $\mathbf{U}^{\mu}, \mathbf{U}^{\sigma},
\mathbf{U}^{\xi}$ and not $\bolds{\tau}$. If values in
$\bolds{\tau}$ are too small, then the variability
in each $\theta^\psi(s)$ is forcibly large and may cause problems in
estimating the diagonal of $\mathbf{T}$ which relates to the variability
of each $\mathbf{U}^{\psi}$. Sensitivity analysis was performed to
ensure these values have little effect on
inference (not shown for conciseness).

The Wishart prior for the precision matrix $\mathbf{T}$ was given
the following mean: $\operatorname{diag}(0.02, 4, 40)'$.
As with $\bolds{\tau}$, these values were calibrated by fitting
independent point process models
and were chosen to reflect the associated levels of variability for
each of $\mu(s)$, $\log(\sigma(s))$ and $\xi(s)$.

\begin{figure}

\includegraphics{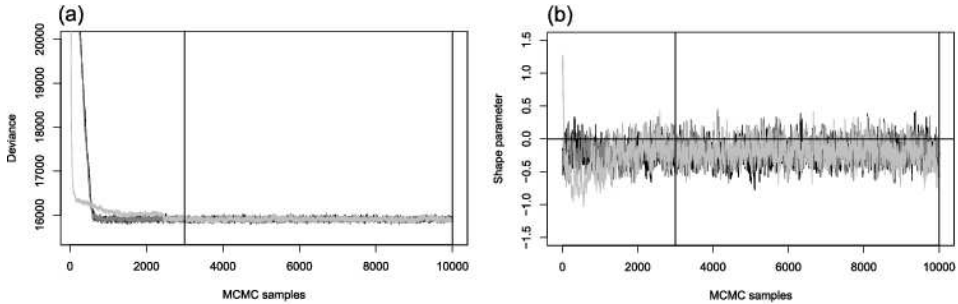}

\caption{\textup{(a)} Deviance samples from each of the three MCMC chains.
Vertical lines denote
the burn-in and the total number of simulations. Samples between the
two lines
are used for inference. \textup{(b)} Samples of the shape parameter $\xi(s)$ for
the grid cell containing London.}
\label{MCMCchains}
\end{figure}

The model in (\ref{cyclonesModel})--(\ref{xi}) was implemented in R
[\citet{R:2012}] using three parallel MCMC chains. These were run on
a workstation with a 3.07~GHz i7 processor and the processing speed for
each chain was 30 seconds for 1000 samples. A total of 50,000
samples were collected per chain and thinned by 5 to reduce
auto-correlation. After thinning, the first 3000 samples from each
chain were
discarded based on a trace plot of deviance (minus twice the
log-likelihood) shown in Figure~\ref{MCMCchains}(a). Convergence in the deviance
is a good indication of convergence to the joint posterior of all
parameters [\citet{Gelman:2004}]. Summarising, 21,000 posterior samples were
used to calculate posterior distribution statistics for the parameters.
Figure~\ref{MCMCchains}(b) shows an example trace plot of $\xi(s)$
for the grid cell containing the London coordinate.

\begin{figure}

\includegraphics{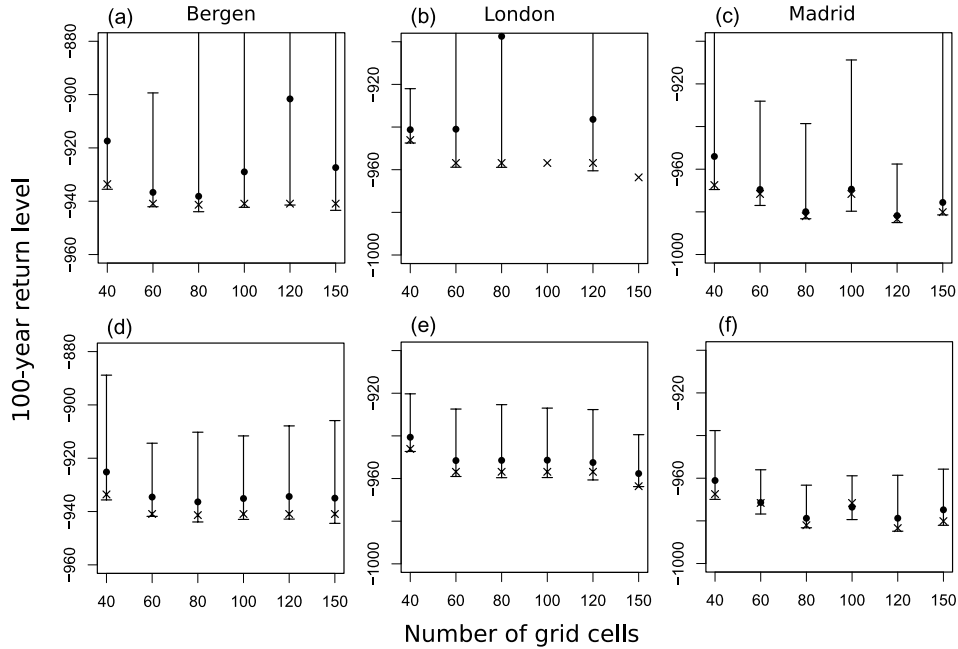}

\caption{Dots are posterior means of the 100-year return level of
$X(s,t)$ versus number of cells in different grid specifications,
along with $95\%$ credible intervals. Left [\textup{(a)} and \textup{(d)}], middle [\textup{(b)} and~\textup{(e)}]
and right [\textup{(c)} and \textup{(f)}] panels refer to the Bergen, London and Madrid
cells, respectively. Top [\textup{(a)}, \textup{(b)}, \textup{(c)}] and bottom [\textup{(d)}, \textup{(e)}, \textup{(f)}] panels refer to
the stationary and the spatial models, respectively. For reference,
the deepest recorded value of $X(s,t)$ in each cell is shown with a
cross symbol.}
\label{convergence}
\end{figure}

\subsection{Sensitivity to grid cell size}\label{Sensitivity}
A purely spatial model [i.e., model (\ref{cyclonesModel})--(\ref{xi})
without $z_1$ and $z_2$] and a stationary model [i.e., model
(\ref{cyclonesModel})--(\ref{xi}) without $z_1$, $z_2$ and the random
effects] were implemented for different grid configurations.
For each model, the 100-year return level
(i.e., the level exceeded by the annual maximum in any particular year
with probability 0.01) of $X(s,t)$ was calculated using (\ref{RL}),
for each of the three coordinates marked in Figure~\ref{MAPS}(c).
Figure~\ref{convergence} shows the posterior mean of the 100-year
return level against the number of cells in each grid configuration
along with $95\%$ credible intervals for each model.
Convergence of the return value, as the number of cells increases, is
evident for the spatial model (although this varies slightly due
sampling variation). The random effects pool information spatially,
whereas the
stationary model ignores neighbouring cells, resulting in failure to
converge, especially over London. Pooling also results in
notably smaller credible intervals for the spatial model---note that
the intervals are skewed. We chose $N=150$ cells for the analysis so
that all cells have an adequate number of nadirs (ranging from 14 to 376).

\section{Results}
Posterior distributions for global parameters are summarised in Table~\ref{posteriors}. Latitude has a positive linear effect on both
the location and log-scale parameters of extreme cyclone depth
$X(s,t)$. The overall NAO effect $\nu$ is positive, in agreement with
findings from previous studies [\citet{Pinto:2009}].
To assess MCMC convergence, the Gelman and Rubin $\hat{R}$ multi-chain\vspace*{1pt}
diagnostic was used for each of our model parameters
[\citet{Gelman:2004}]. The $\hat{R}$ values for each parameter in
Table~\ref{posteriors} are all close to unity, suggesting convergence.

\begin{figure}

\includegraphics{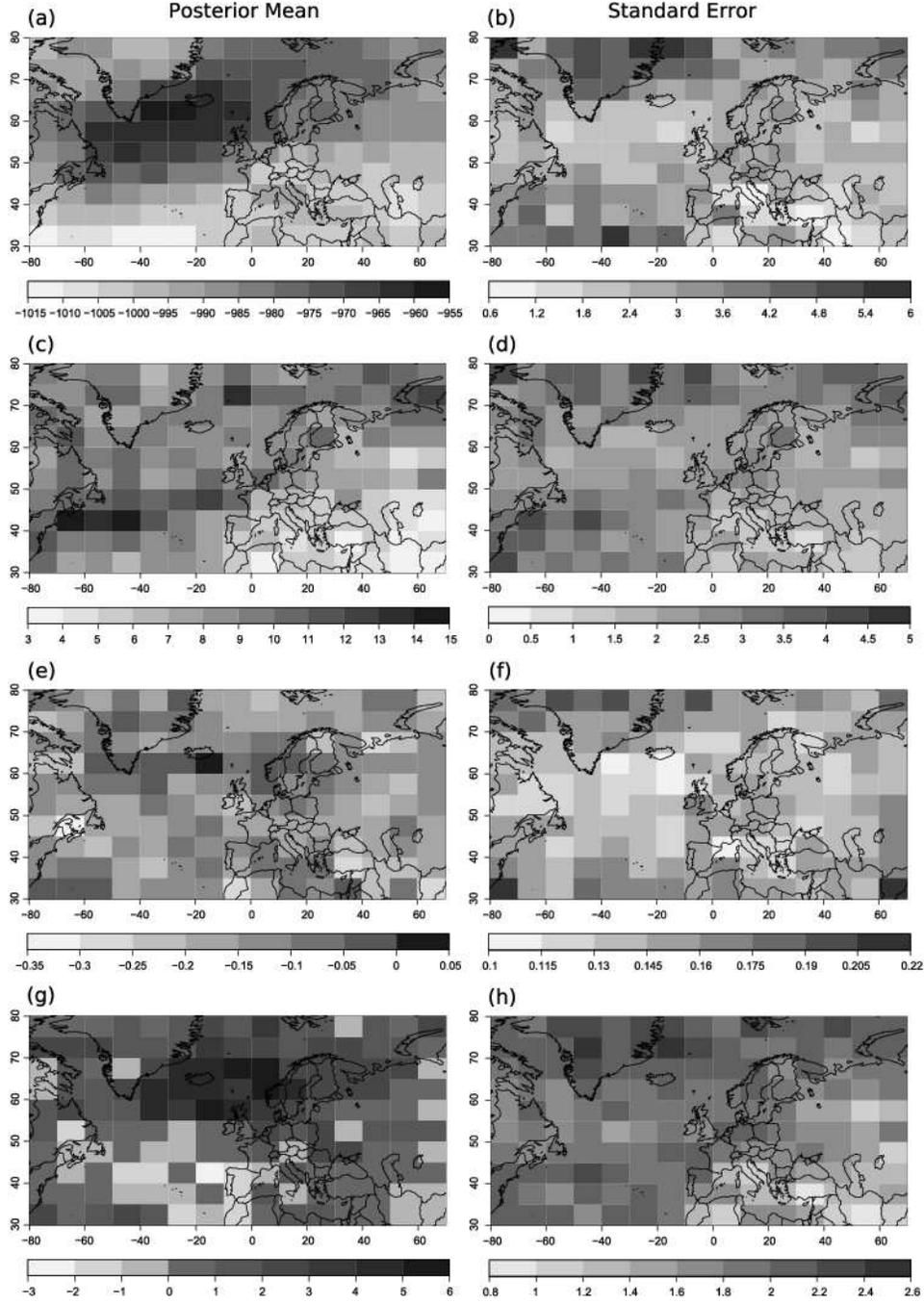}

\caption{Posterior means for \textup{(a)} $\mu(s,t)$, \textup{(c)} $\sigma(s,t)$, \textup{(e)} $\xi
(s)$ and \textup{(g)} $\beta_2(s)$
and standard errors in~\textup{(b)}, \textup{(d)}, \textup{(f)} and \textup{(h)}, respectively, where
$z_1(t)$ is latitude at centre of grid cell
and $z_2(t)=0$.}
\label{MuSigXi}
\end{figure}

Figure~\ref{MuSigXi} shows posterior means and standard deviations of
$\mu(s,t)$, $\sigma(s,t)$ and $\xi(s)$.
Much of the spatial structure in the extreme nadirs comes from the
location and scale parameters. The
posterior means for the shape parameter $\xi(s)$ are more uniform and
generally negative, apart from one cell over Iceland. Exploring
this further, the two deepest nadirs in the reanalysis occurred in this
cell, and they are considerably lower than the rest of the nadirs
in the vicinity. A return level plot from the particular cell indicated
that the two nadirs (one of them being from the record-breaking Braer
cyclone) unduly influenced the sign of the shape parameter. This has
been quantified by removing those two
points and refitting the model, however, this being an analysis of
extremes, it makes little sense to remove such values.

\begin{table}
\tabcolsep=0pt
\caption{Summary of parameter posterior distributions}\label{posteriors}
\begin{tabular*}{\textwidth}{@{\extracolsep{\fill}}lcd{4.8}cc@{}}
\hline
\textbf{Parameter} & \textbf{Prior} & \multicolumn{1}{c}{\textbf{Posterior mean (s.e.)}} & $\bolds{95\%}$ \textbf{Cr.I.} & \multicolumn{1}{c@{}}{$\bolds{\hat{R}}$}\\
\hline
$\beta^{\mu}_1$ (Latitude)& $\mathrm{N} (0,100)$ & 4.71\
(0.62)&$[3.61,5.94]$ & 1.13\\[2pt]
$\beta^{\sigma}_1$ (Latitude)& $\mathrm{N} (0,100)$ &0.12\
(0.07)&$[0.00,0.25]$ & 1.03\\
Overall NAO effect $\nu$ & $\mathrm{N} (0,100)$ & 1.21\
(0.24)&$[0.77,1.66]$& 1.01\\
Variance NAO effect $\phi^2$ & $\propto1/\phi^2$ &5.6\
(1.85)&$[3.09,10.25]$& 1.00 \\
$\beta^{\mu}_0$ & $\mathrm{N} (-944.1,100)$ & -987.4\
(0.51)&$[-988.5,-986.5]$ & 1.04 \\[2pt]
$\beta^{\sigma}_0$ & $\mathrm{N} (5.7,100)$ & 2.03\ (0.06)&$[1.91,2.15]$ &
1.05\\[2pt]
$\beta^{\xi}_0$ & $\mathrm{N} (-0.19,100)$ & -0.13\ (0.03)&$[-0.18,-0.07]$&
1.10\\
\hline
\end{tabular*}
\end{table}

\begin{figure}

\includegraphics{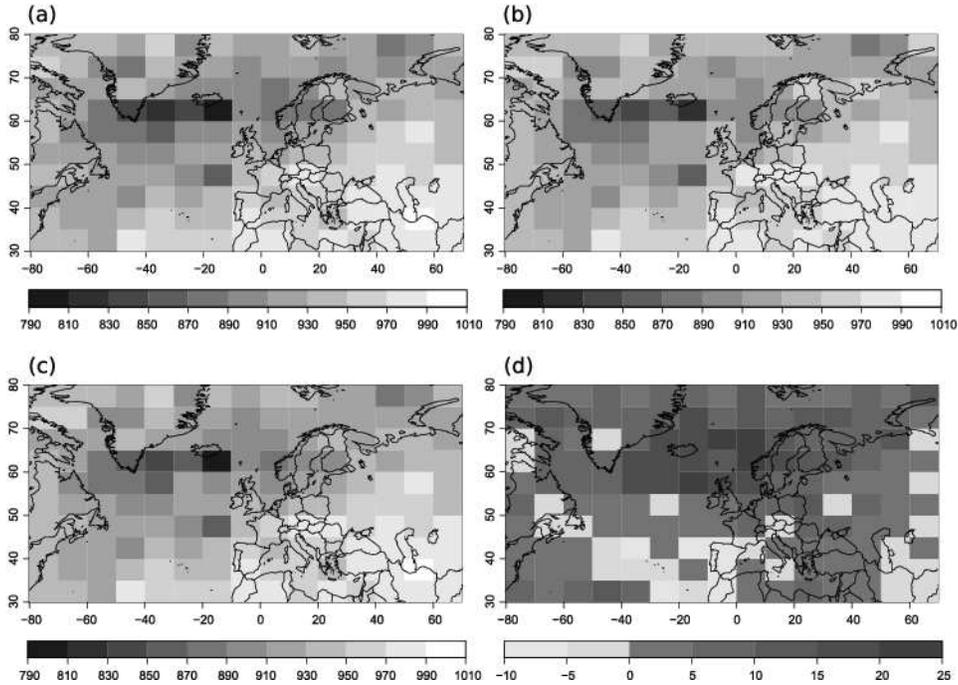}

\caption{Estimated lower limits of nadir sea-level pressure for \textup{(a)}
$\operatorname{NAO} = 2$, \textup{(b)} $\operatorname{NAO} = -2$
and \textup{(c)}~$\operatorname{NAO} = 0$. \textup{(d)} shows the difference between \textup{(a)} and \textup{(b)}.}
\label{NAOub}
\end{figure}

A negative shape parameter implies that the distribution of extreme
cyclone depth $X(s,t)$, at time $t$ and cell $s$,
has an upper bound given by $\sigma(s,t)/\xi(s)-\mu(s,t)$. Here this
corresponds to a lower limit on nadir
sea-level pressure. Many of the posteriors for $\xi(s)$ do have some
mass over the positive real line [see, e.g., Figure~\ref{MCMCchains}(b)]. However, except for the Iceland cell, the negative
masses for $\xi(s)$ are all greater than 0.5, therefore, we
can use the negative posterior $\xi(s)$ samples to obtain a conditional
posterior distribution for the estimated lower limit.
The posterior means of these limits are shown in Figure~\ref{NAOub}(c)
for $\operatorname{NAO} = 0$. The limit for the cell containing Bergen is
890.0~hPa [193.0, 932.6]\vadjust{\goodbreak} and for the London cell it is 943.0~hPa [714.8,
959.4], whereas in the Madrid cell it is 953.5~hPa [537.9, 978.7].
The $95\%$ credible intervals are skewed and noticeably wide, which is
to be
expected given we are trying to estimate the 100th percentile. The
lower bounds on some of
these intervals are too low to be physically plausible and this
reflects the fact that the statistical model is not constrained by
physical mechanisms. Note also that there is considerable literature
focusing on the problem of estimating upper/lower bounds of
distributions. See \citeauthor{deHaan:2006} [(\citeyear{deHaan:2006}), Chapter~4] for a detailed
discussion and a description of both maximum likelihood and moment
estimators for bounds arising from extreme value distributions. In
addition, \citet{Einmahl:2008} provide refined estimators for
bounds of world records in athletics and their respective sampling
distributions.

The posterior means and standard deviations of the NAO effects $\beta
_2(s)$ are shown in Figure~\ref{MuSigXi}(g) and
(h), respectively. A positive effect is prominent in the area
where cyclones deepen the most: in the vicinity of Iceland, northern
Europe and Scandinavia. A negative effect is also apparent, effectively
over Spain and the Azores. This North--South NAO effect in the central
Atlantic is consistent with the exploratory diagnostics in Figure~\ref{NAOplots}(c) and (d). Maps of the estimated lower limit for
$\operatorname{NAO} = -2$ and $\operatorname{NAO} = 2$ are given in Figure~\ref{NAOub}(a) and (b). To better see the effect of NAO on the
estimated lower
limit, Figure~\ref{NAOub}(d) shows the difference in hPa between the
estimated lower limits for $\operatorname{NAO} = 2$\vadjust{\goodbreak} and $\operatorname{NAO} = -2$. The
difference can get up to 25~hPa in the area where NAO has the biggest
effect, that is, northern Europe and Scandinavia.

\begin{figure}

\includegraphics{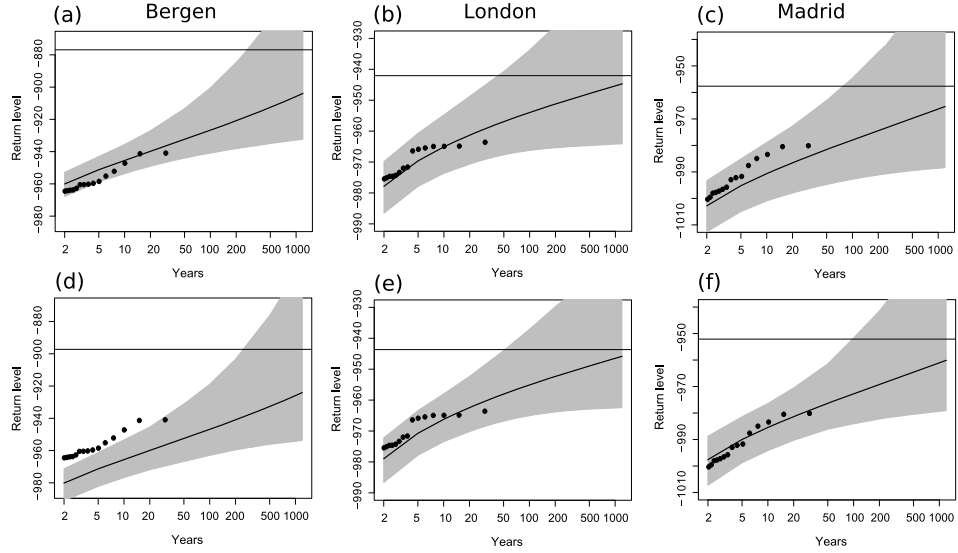}

\caption{Individual grid cell return level plots (posterior means) with
$95\%$ credible intervals.
Observed values shown in solid circles. Top panel: $\operatorname{NAO} = 2$;
bottom panel: $\operatorname{NAO} = -2$.
Left panel: Bergen cell; middle panel:
London cell; right panel: Madrid cell. Horizontal lines are estimated
upper bounds of $X(s,t)$ for
$\operatorname{NAO} = 2$ (top) and $\operatorname{NAO} = -2$ (bottom).}
\label{RL_plots}\vspace*{-3pt}
\end{figure}

Figure~\ref{RL_plots} shows return level plots of $X(s,t)$ for the
Bergen--London--Madrid grid cells, for $\operatorname{NAO} = \pm2$.
Note that this is not a goodness-of-fit test, as each point in these
plots (the recorded
value) is associated with a different NAO value, whereas the return
level curves are calculated at $\operatorname{NAO} = \pm2$.
A positive/negative NAO effect is noticeable in the Bergen/Madrid
cells, confirming the NAO North--South effect.
No NAO effect is evident in the London cell. The horizontal line in
each plot is the estimated cyclone depth limit, suggesting that
for all three cells, nadirs could have been much deeper than the ones recorded.

\begin{figure}

\includegraphics{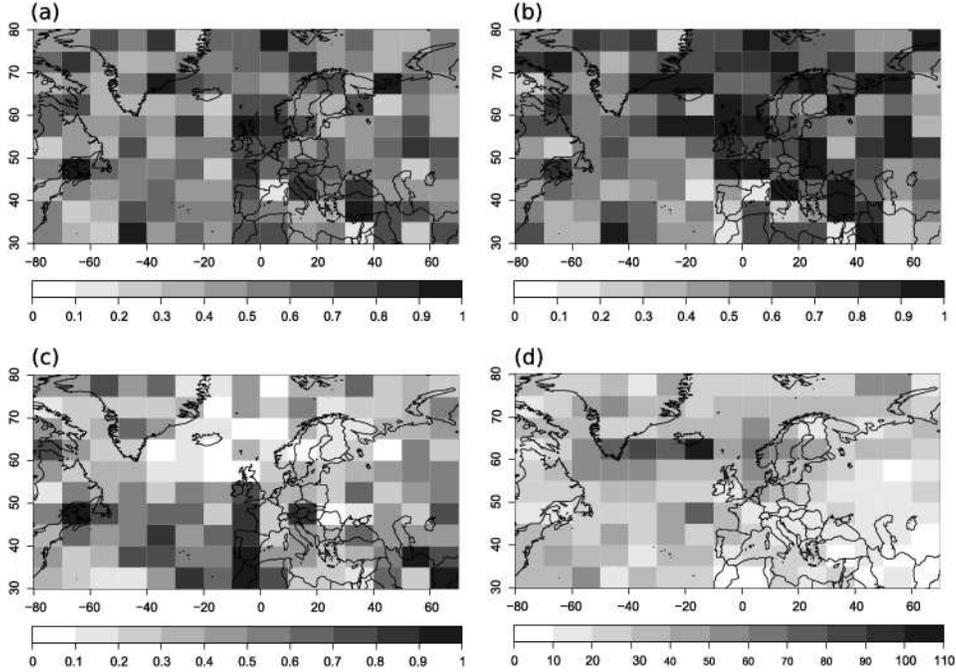}

\caption{Probability of observing a deeper nadir than the recorded
30-year deepest nadir in each cell:
\textup{(a)} Calculated for NAO values associated with the recorded values
nadirs, \textup{(b)} $\operatorname{NAO} = 2$
and \textup{(c)}~$\operatorname{NAO} = -2$. \textup{(d)} The difference in hPa between the
estimated depth limit and the deepest recorded 30-year
nadirs in each cell.}\vspace*{-6pt}
\label{ProbMax}
\end{figure}

Therefore, we also consider the quantity $\pi(s,t)=\operatorname{Pr} (X(s,t)
> x_{m}(s) )$, where $x_{m}(s)$ is the negated minimum recorded nadir
in grid cell $s$ for the 30-year period. (Note that this is equivalent
to describing how unusual the recorded
depth was, rather than the probability of ever getting deeper than the
recorded 30-year minimum nadir.)
We transform the GEV parameters to reflect the distribution of 30-year,
rather than yearly depth values:
$\tilde{\sigma}=\sigma\delta^\xi$ and $\tilde{\mu} = \mu+ \tilde{\sigma
}(1-\delta^{-\xi})/\xi$ where
$\delta=30$. Figure~\ref{ProbMax}(a) shows $\pi(s,t)$ for values of NAO
associated with $x_{m}(s)$.
There are high values of $\pi(s,t)$, especially over western Europe.
Figure~\ref{ProbMax}(b) shows $\pi(s,t)$ for $\operatorname{NAO} = 2$, indicating
that for a positive NAO phase there is high
probability of deeper nadirs over Europe, Iceland and Scandinavia. For
$\operatorname{NAO} = -2$, Figure~\ref{ProbMax}(c)\vadjust{\goodbreak}
attributes high probability of deeper nadirs over Spain, Portugal, west
of France and also over the Azores region.
Furthermore, Figure~\ref{ProbMax}(d) shows the difference in hPa between
the estimated depth limit for MSLP
[Figure~\ref{ProbMax}(a)] and $x_m(s)$ for each cell. For most cells, the
difference is in the
range of 10--50~hPa, while for cells over Iceland the range is
80--110~hPa, indicating the 30-year reanalysis
is not long enough to capture nadir depths near the estimated limits.

\begin{figure}

\includegraphics{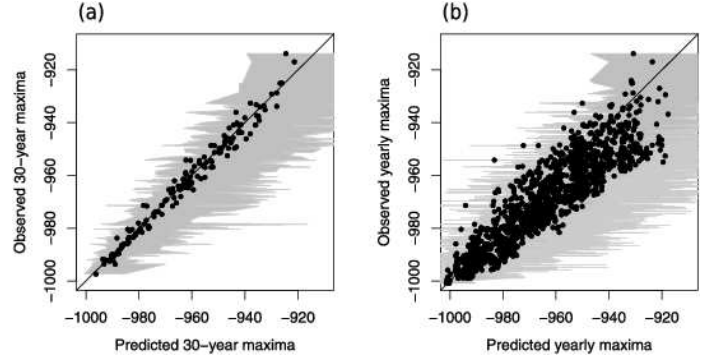}

\caption{Recorded versus predicted values of: \textup{(a)} 30-year deepest
nadirs in each cell and \textup{(b)}~yearly deepest nadirs
in each cell. The predicted values are the means of the posterior
predictive distributions while the grey shaded area
represents the associated $95\%$ prediction intervals.}
\label{maxima}
\end{figure}

We use posterior predictive checking \citeauthor{Gelman:2004} [(\citeyear{Gelman:2004}), Chapter~6] to
assess model fit. This compares each observation, $x(s,t)_{\mathrm{obs}}$, to the
posterior predictive distribution for replications, $X(s,t)_{\mathrm{rep}}$, of
$X(s,t)$ given the data, $D$, used to fit the model. If the observations
do not behave as if they are sampled from their posterior predictive
distributions, then this indicates poor model fit. Samples of
$X(s,t)_{\mathrm{rep}}$ were obtained by simulating from GEV distributions with
parameters equal to draws from their joint posterior distribution and
then the posterior predictive means and $95\%$ posterior predictive
intervals were approximated from these samples. We plot the
observations of (a) the deepest 30-year nadirs and (b)~the deepest
yearly nadirs against the corresponding posterior predictive means and
intervals in Figure~\ref{maxima}(a) and (b), respectively. None of the
observations seem extreme with respect to the posterior predictive
distributions: the 45-degree line falls well within the prediction intervals.\vadjust{\goodbreak}

\begin{figure}[b]

\includegraphics{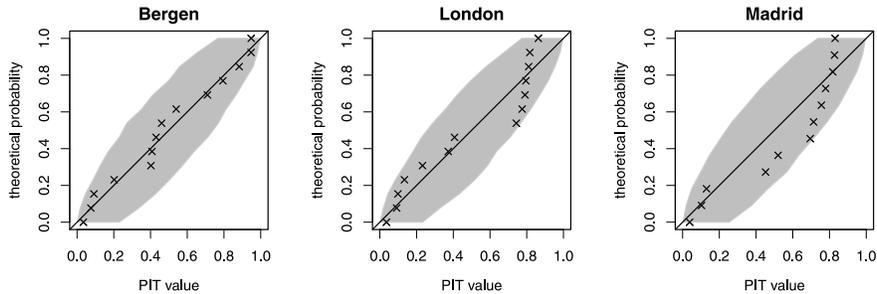}

\caption{Probability--probability plots of theoretical $\operatorname
{Unif}(0,1)$ probabilities versus probability integral transform (PIT)
values $z(s,t)$ for the Bergen, London and Madrid cells. The $95\%$
confidence intervals reflect sampling uncertainty.}
\label{PPplots}
\end{figure}

We also calculate the probability integral transform (PIT), $z(s,t)= \break \Pr
 (X(s,t)_{\mathrm{rep}}\leq x(s,t)_{\mathrm{obs}}\mid D )$, of each observation
relative to its posterior predictive distribution. If the model is a
good fit, then the $z(s,t)$ should follow a uniform distribution on the
interval $(0,1)$. For each grid cell, $s$, we plot the probability
points $(i-1)/(n(s)-1)$ for $i=1,\ldots,n(s)$ against the order
statistics of the $z(s,t)$ values for that cell, where $n(s)$ is the
number of observations in cell~$s$. Departures from the 45-degree
line indicate poor model fit. We indicate the sampling variation that
would be expected in these
plots when the model is perfect by pointwise $95\%$ confidence
intervals, constructed by simulating samples of size $n(s)$ from the
uniform distribution on $(0,1)$. Figure~\ref{PPplots} shows these plots
for Bergen, London and Madrid. No points fall outside the $95\%$
intervals, indicating adequate fit. Note that PIT values are often used
in forecast verification; see, for instance, \citet{Gneiting:2007}
and references therein. Although histograms are the more conventional
way of displaying PIT values, here we only have a few data points
for each cell, so we use probability--probability plots.

\section{Conclusions}
We have implemented a flexible model, adapted from \citet{Cooley:2008},
to reanalysis cyclone data in what we believe to
be the first study that simultaneously models both the spatial and
temporal structure of extreme extra-tropical cyclones.
Using (1) spatial random effects, (2) latitude as a covariate and (3) a
150 cell spatial
regularisation, spatial variation was adequately modelled in the
extremal behaviour of the cyclones.
The North Atlantic Oscillation was used as a covariate and was found to
have a
significant effect on extremal cyclone behaviour, especially over
Northern Europe and the Iberian peninsula.

Although this is a first step toward studying the spatio-temporal
behaviour of extreme cyclones, the analysis relies on assumptions
which may oversimplify the problem: (1) the creation of an artificial
grid, (2) the choice of threshold in each cell and (3) the subjective
choice of spatial proximity. The choice of the grid is a potential
weakness which can introduce bias, as both the number of cells and their
shape are subjectively chosen. Techniques such as Dirichlet
tessellation or Delaunay triangulation [\citet{Illian:2008}] may be useful
for defining a more optimal ``data-driven'' grid. The shape of the cells
is particularly important if one is interested in modelling
data along cyclone tracks rather than individual points as in our
application. If interest was in the relative spatial cyclone impact,
one could use cell-specific rather than cyclone-specific nadirs,
rendering the rectangular cells inappropriate. Hexagonal cells
would be more appropriate as illustrated in an application to tropical
cyclones in \citet{Elsner:2012}.
Threshold choice in each cell may also prove to be an issue. Ideally,
model fit should be one of the criteria for choosing the threshold.
For the application in this paper, three different thresholds were
considered: the $85\%$, $90\%$ and $95\%$ quantile in each cell.
Model fit diagnostics (Figures~\ref{maxima} and \ref{PPplots})
indicated worse fit for the $85\%$ quantile, and an identical fit for the
higher quantiles---which is why we selected the $90\%$ quantile for
model implementation.
To avoid choosing the threshold altogether, one might instead estimate
the threshold from the data.
For example, we have explored the possibility of using a mixture model
as in \citet{Frigessi:2002} where the threshold is estimated
but which also allows all available data (not just the extremes) to be
used for each cell, which in turn allows the use of a finer grid.
Last, the proximity structure used to define the covariance matrix of
the spatial random effects is also an assumption which can
affect the degree of spatial smoothing. The spatially random occurrence
of cyclone nadirs was ``marginalised'' here by dividing the
region into grid cells, whereas one should ideally try to model both
the spatial occurrence and intensity at the same time, for example, by
using spatial marked point process models. Nevertheless, despite these
assumptions, the model in this study is flexible enough to
be used in other similar studies, for example, ones involving tropical
cyclone wind speed maxima or cyclone-related peak precipitation.

\begin{appendix}\label{app}
\section*{Appendix}
\subsection{Point process model for extremes}\label{Appendix}
The point process model for extremes involves a
bivariate variable $Y=(X,T)$, with $T \in[0,1]$ being a scaled random
variable associated with time and $X \in\mathbb{R}$ a random
variable associated with intensity. The model is a marked point
process, which for $X>u$ (a high threshold) under some
linear normalisation and mixing criteria [\citet{Smith:1989}] behaves
like a nonhomogeneous Poisson process with intensity function
%
\begin{equation}
\label{intensity} \lambda(x,t) = \frac{1}{\sigma} \biggl[ 1+\xi \biggl(
\frac{x-\mu}{\sigma} \biggr) \biggr]^{-1/\xi- 1},
\end{equation}
provided that $1+(\xi/\sigma)(x-\mu) > 0$. The intensity function
$\lambda(x,t)$ is zero for $1+(\xi/\sigma)(x-\mu) < 0$.
The exceedance rate is explicitly modelled in terms of the mean number
of exceedances in the time interval $[t_1,t_2]$:
\[
\Lambda\bigl([t_1,t_2]\times(u,\infty)\bigr) =
(t_2 - t_1) \biggl[1+\xi \biggl(\frac
{u-\mu}{\sigma} \biggr)
\biggr]^{-1/\xi}.
\]

The likelihood given observations $y_i=(x_i,t_i)$ in region
$[0,1]\times(u,\infty)$ is
%
\begin{eqnarray}
L(\mu,\sigma,\xi;\mathbf{x},\mathbf{t}) &=& \exp \biggl\{-n_y \int
_0^1 \int_u^{\infty}
\lambda(x,t) \,dx \,dt \biggr\} \prod_i \lambda
(x_i,t_i) \label{Integral}
\\
&= & \exp \biggl\{-n_y \biggl[ 1+\xi \biggl( \frac{u-\mu}{\sigma}
\biggr) \biggr]^{-1/\xi} \biggr\} \prod_i
\lambda(x_i,t_i), \label{PP_lik}
\end{eqnarray}
where $n_y$ is the number of years of observed data so that parameters
$\mu$, $\sigma$ and $\xi$ correspond to the GEV distribution of
yearly maxima. Because the time variable $T$ does not actually appear
in (\ref{intensity}) and thus in (\ref{PP_lik}), we use the
concise notation $X\sim\operatorname{PP}(\mu,\sigma,\xi,u)$ as in Section~\ref{model_spec}.
The likelihood contribution from a single event $(x_i,t_i)$ is
\[
\mathrm{L}(\mu,\sigma,\xi,u) = \exp \biggl\{-n_y[t_i-t_{i-1}]
\biggl[ 1+\xi \biggl( \frac{u-\mu}{\sigma} \biggr) \biggr]^{-1/\xi} \biggr\}
\lambda(x_i,t_i),
\]
for $i=0,\ldots,n$ where $n$ is the number of events. Note that $t_0=0$
and that the likelihood contribution,
for the time interval between the last event occurrence and $t=1$, is
the probability of no events in the interval, that is,
\[
\exp \biggl\{-n_y[1-t_n] \biggl[ 1+\xi \biggl(
\frac{u-\mu}{\sigma} \biggr) \biggr]^{-1/\xi} \biggr\}.
\]
%

The conditional model in (\ref{cyclonesModel})--(\ref{xi}) was
implemented using the likelihood (\ref{PP_lik}) for each cell. However,
because of the temporal covariates, the outermost integral over time in
(\ref{Integral}) is impossible to calculate analytically
unless one knows explicitly how the covariates evolve in time.\vadjust{\goodbreak} A remedy
is to approximate the integral: divide the time range in small
intervals with endpoints $0=k_1,k_2,\ldots,k_J=1$ and assume the
function is constant in each interval. The integral
\[
\int_0^1 \biggl[ 1+\xi(s) \biggl(
\frac{u(s)-\mu(s,t)}{\sigma(s,t)} \biggr) \biggr]^{{-1}/{\xi(s)}} \,dt
\]
is thus approximated by the Riemann sum
\[
\frac{1}{J}\sum_{i=1}^{J} \biggl[ 1+
\xi(s) \biggl( \frac{u(s)-\mu
(s,k_i)}{\sigma(s,k_i)} \biggr) \biggr]^{{-1}/{\xi(s)}},
\]
where $J$ is the number of intervals. In practice, $J$ is determined by
observations of the covariates for all data (not just the extremes).

\subsection{Measures of extremal dependence}\label{Extremal_dependence}
The measure of extremal dependence $0< \chi< 1$ between random
variables $Z$ and $W$ is defined as
\[
\chi= \lim_{p\rightarrow1} \operatorname{Pr}\bigl(F_{Z}(Z) >
p | F_{W}(W) > p\bigr) = \lim_{p \to1} \chi(p),
\]
where $F_Z$ and $F_W$ are the respective distribution functions of $Z$
and $W$. The other extremal dependence measure
$-1\leq\bar{\chi} \leq1$ is defined as
\[
\bar{\chi} = \lim_{p\rightarrow1} \frac{2\log\operatorname{Pr} (F_Z(z) >
p )}{\log\operatorname{Pr} (F_Z(z)>p,F_W(w)>p )} - 1 = \lim
_{p\to1} \bar{\chi}(p).
\]
If $\chi> 0$ and $\bar{\chi}=1$, the two variables are asymptotically
dependent and $\chi$ measures the strength of that dependence.
If $\chi= 0$ and $\bar{\chi}<1$, the two variables are asymptotically
independent, in which case $\bar{\chi}$ measures the strength
of dependence---within the class of asymptotically independent variables.
Roughly, $\bar{\chi}$ measures the ``speed'' at which $\chi(p)$
approaches zero.
\citet{Coles:1999} advocate the use of both $\chi$ and $\bar{\chi}$ as
indicators of extremal
dependence, providing complementary information on different aspects of
that dependence. 

\subsection{Intrinsic AutoRegressive priors}\label{IAR}
Consider a grid with $N$ cells. If the random effect $\bolds{\phi
}=(\phi(1),\ldots,\phi(N))'$ is assumed to have an IAR
prior, then $\bolds{\phi} \sim N(0,(\tau\mathbf{W})^{-1})$,
where $\mathbf{W}$ is the adjacency matrix and
the conditional distribution for each $\phi(s)$ given the rest is given by
\[
\phi(s)|\bolds{\phi}(-s) \sim N \biggl(\bar{\phi}(s),\frac{1}{\tau
m(s)} \biggr),
\]
where $\bolds{\phi}(-s)$ is $\bolds{\phi}$ excluding $\phi
(s)$; $\bar{\phi}(s)$ is the average of $\bolds{\phi}(-s)$
that are adjacent to $\phi(s)$ and $m(s)$ is the number of those adjacencies.
\end{appendix}

\section*{Acknowledgements}
We kindly thank Dan Cooley for providing the R code which was adapted
for this study and for insightful comments. We
also thank Kevin Hodges for providing the track data for NCEP-CFS.

%





\printaddresses
\end{document}